\documentclass[12pt]{article}
\usepackage{amstext,amsthm,amsmath,amssymb,amscd,amsfonts,latexsym}
\usepackage[dvips]{graphicx}
\usepackage{graphicx,epsfig}
\usepackage{amsmath,amssymb,amstext, listings, multirow,epsfig, graphicx,color,subfig}
\usepackage{appendix}
\usepackage{units}
\usepackage{natbib}
\usepackage{tabularx}
\usepackage{caption}
\makeatletter
\def\hlinewd#1{%
\noalign{\ifnum0=`}\fi\hrule \@height #1 %
\futurelet\reserved@a\@xhline}
\makeatother

\newcommand{\be}{\begin{enumerate}}
\newcommand{\ee}{\end{enumerate}}

\theoremstyle{plain}

\newtheorem{theorem}{\bf Theorem}[section]

\newtheorem{defn}{\bf Definition}[section]

\defcitealias{BCBS}{BCBS [2006]}
\defcitealias{Basel3}{BCBS [2011]}
\defcitealias{CRMPG}{CRMPG [2005]}
\defcitealias{TradingBookReview}{BCBS [2012]}

\begin{document}
\title{Wrong-Way Bounds in Counterparty Credit Risk Management
\thanks{The authors are grateful to John Chadam, Satish Iyengar, Dan Rosen, Lung Kwan Tsui, and participants at the Second Qu{\'{e}}bec-Ontario
Workshop on Insurance Mathematics and the 2012 Annual Meeting of the Canadian Applied and Industrial Mathematics Society for many interesting discussions and
helpful comments.}}
\date{July 7, 2014}
\vspace{0.5in}

\author{Amir Memartoluie\thanks{Corresponding author. David R. Cheriton School of Computer Science, University of Waterloo.
amemartoluie@uwaterloo.ca}
\\ David Saunders \thanks{Department of Statistics and Actuarial Science, University of Waterloo. dsaunders@uwaterloo.ca. Support from an NSERC Discovery Grant is
gratefully acknowledged.}
\\ Tony Wirjanto \thanks{School of Accounting and Finance, and Department of Statistics and Actuarial Science, University of Waterloo. twirjant@uwaterloo.ca}}
\maketitle
\begin{abstract}
 We study the problem of finding the worst-case joint distribution of a set of risk factors given prescribed multivariate marginals and a nonlinear loss function. We show that when the risk measure is CVaR, and the distributions are discretized, the problem can be conveniently solved using linear programming technique. The method has applications to any situation where marginals are provided, and bounds need to be determined on total portfolio risk. This arises in many financial contexts, including pricing and risk management of exotic options, analysis of structured finance instruments, and aggregation of portfolio risk across risk types. Applications to counterparty credit risk are emphasized, and they include assessing wrong-way risk in the credit valuation adjustment, and counterparty credit risk measurement. Lastly a detailed application of the algorithm for counterparty risk measurement to a real portfolio case is also presented in this paper.
\end{abstract}

\thispagestyle{empty}
\newpage

\begin{section}{Introduction}
 In recent years counterparty credit risk management has become an increasingly important topic for both regulators and participants in over-the-counter
 derivatives markets. Even before the financial crisis, the Counterparty Risk Management Policy Group noted that counterparty risk is ``probably the single most important variable in  determining whether and with what speed financial disturbances become financial shocks, with potential systemic traits'' (\citetalias{CRMPG}). This concern over counterparty credit risk as a source of systemic stress has been reflected in the historical developments in the Basel Capital Accords (\citetalias{BCBS}, \citetalias{Basel3}, see also Section 3 below).

 Counterparty Credit Risk (CCR) is defined as the risk of loss due to default or the change in creditworthiness of a counterparty before the final settlement of the cash flows of a contract. An examination of the problems of measuring and managing this risk reveals a number of key features. First, the risk is bilateral, and current exposure can lie either with the institution or its counterparty. Second, the evaluation of exposure must be done at the portfolio level, and must take into account relevant credit mitigation arrangements, such as netting and the posting of collateral, which may be in place. Third, the exposure is stochastic in nature, and contingent on current market risk factors, as well as the creditworthiness of the counterparty, and credit mitigation. In addition, possible dependence between credit risk and exposure, known as wrong-way risk, is also an important modelling consideration.\footnote{Since in general the risk is bilateral, in the case of pricing contracts subject to counterparty credit risk (i.e. computing the {\em credit valuation adjustment\/}), the creditworthiness of both parties to the contract is relevant. In this paper, we take a unilateral perspective, focusing exclusively on the creditworthiness of the counterparty.} Finally, the problem is computationally highly intractable. To calculate risk measures for counterparty credit risk, one requires the {\em joint distribution\/} of all market risk factors affecting the portfolio of (possibly tens of thousands of) contracts with the counterparty, as well as the creditworthiness of both counterparties, and values of collateral instruments posted. It is nearly impossible to estimate this joint distribution accurately. Consequently, we are faced with a problem of risk management under {\em uncertainty}, where at least part of the probability distribution needed to evaluate the risk measure is unknown.

 Fortunately, we do have partial information to aid in the calculation of counterparty credit risk. Most financial institutions have in place models for simulating the joint distribution of counterparty exposures, created (for example) for the purpose of enforcing exposure limits. Additionally, internal models for assessing default probabilities, and credit models (both internal and regulatory) for assessing the joint distribution of counterparty defaults are also available at our disposal. We can view the situation as one where we are given the (multi-dimensional) marginal distributions of certain risk factors, and need to evaluate portfolio risk for a loss variable that depends on their joint distribution. The Basel Accord (\citetalias{BCBS}) has employed a simple adjustment based on the ``alpha multiplier'' to address this problem. A stress-testing approach, employing different copulas and financially relevant ``directions'' for dependence between the market and credit factors is presented in \citet{GarciaCespedes} and~\citet{Rosen-Saunders-2009}. This method allows for a computationally efficient evaluation of counterparty credit risk, as it leverages pre-computed portfolio exposure simulations.\footnote{Generally speaking, the computational cost of the algorithms is dominated by the time required for evaluating portfolio exposures - which involves pricing thousands of derivative contracts under at least a few thousand scenarios at multiple time points - rather than from the simulation of portfolio credit risk models.}

 In this paper we investigate the problem of determining the {\em worst-case joint distribution\/}, i.e. the distribution that has the given marginals, and produces the highest risk measure.\footnote{Since the marginals are specified, the problem is equivalent to finding the worst-case joint distribution with the prescribed (possibly multi-variate) marginals.} This approach is motivated by a desire to have conservative measures of risk, as well as to provide a standard of comparison against which other methods be evaluated. While in this paper we focus on the application to counterparty credit risk, as mentioned earlier, the problem formulation is completely general, and can be applied to other situations in which marginals for risk factors are known, but the joint distribution is unknown. Finally, we note that we work with Conditional Value-at-Risk (CVaR), rather than Value-at-Risk (VaR), which is the risk measure that currently determines regulatory capital charges for counterparty credit risk in the Basel Accords (\citetalias{BCBS}, \citetalias{Basel3}). The motivation for this choice is twofold. First, it yields a computationally more tractable optimization problem for the worst-case joint distribution, which can be solved using a linear programming technique. Although not directly in the context of CCR, the Basel Committee is also considering the replacement of  VaR with CVaR as the risk measure for determining capital requirements for the trading book
(\citetalias{TradingBookReview}).

 Model uncertainty, and problems with given marginal distributions or partial information have been studied in many financial contexts. One example is the pricing of
 exotic options, where no-arbitrage bounds may be derived based on observed prices of liquid instruments. Related studies include \citet{Bertsimas-2002}, \citet{Hobson-2005}, \citet{Hobson-Laurence-2005}, \citet{Laurence-Wang-2004}, \citet{Laurence-Wang-2005}, and \citet{Chen-2008}, for exotic options written on multiple assets $(S_1,\ldots,S_T)$ observed at the same time $T$. The approach closest to the one we take in this paper is that of \citet{Mathias-2011}, in which the marginals $(\Psi(S^T_1),\ldots,\Psi(S^T_k))$ are assumed to be given, and an infinite-dimensional linear programming technique is employed to derive price bounds. There is also a large literature on deriving bounds for joint distributions with given marginals, and corresponding  VaR bounds. For a recent survey, see \citet{PuccettiRuschendorf}.  The problems considered in this paper can be characterized by two important aspects; (i) we use an alternative risk measure (CVaR), which are provided with multivariate (non-overlapping) marginal distributions, and (ii) we have losses that are a non-linear (and non-standard) function of the underlying risk factors. \cite{glasserman2014robust} present a similar empirical approach to estimating worst-case error in a range of risk management problems; in addition to discussing the theoretical aspects of this problem, one of the primary goals of our paper is addressing the numerical challenges arising from worst-case joint distribution problem.

 \citet{Haase-Ilg-Werner-2010} propose a model-free method for a bilateral credit valuation adjustment; in particular their proposed approach does not rely on any specific model for the joint evolution of the underlying risk factors.  \citet{Talay-Zhang-2002} treat model risk as a stochastic differential game between the trader and the market, and prove that the corresponding value function is the viscosity solution of the corresponding Isaacs equation. \citet{Avellaneda-1995}, \citet{Denis-Hu-Peng-2010} and \citet{Denis-Martini-2006} consider pricing under model uncertainty in a diffusion context. Recent works on risk measures under model uncertainty include \citet{Kervarec-2008} and \citet{Bion-Nadal-Kervarec-2010}.

 The remainder of the paper is structured as follows. Section 2 frames the problem of finding worst-case joint distributions for risk factors with given marginals, and show how this can be reduced to a linear programming problem when the risk measure is given by CVaR and the distributions are treated as discrete. Section 3 outlines the application of this general approach to counterparty credit risk in the context of the model underlying the CCR capital charge in the Basel Accord. In section 4 a nontrivially numerical example using a real portfolio is provided, and section 5 presents conclusions and directions for future research.
\end{section} \\ \\ \\

\section{Worst-Case Joint Distribution Problem}
 Let $Y$ and $Z$ be two vectors of risk factors. We assume that the multi-dimensional marginals of $Y$ and $Z$, denoted by $F_{Y}(\overrightarrow{y})$ and $F_{Z}(\overrightarrow{z})$ respectively, are known, but that the joint distribution of $(Y,Z)$ is unknown (Note: in the context of counterparty credit risk management discussed in the next section, $Y$ and $Z$ will be vectors of market and systematic credit factors  respectively). Portfolio losses are defined to be $L= L(Y,Z)$, where in general this function can be non-linear. We are interested in determining the joint distribution of $(Y,Z)$ that maximizes a given risk measure $\rho$:
 \begin{equation}\label{WCR}
  \max_{\mathfrak{F}(F_Y,F_Z)} \rho\left( L(Y,Z)\right)
 \end{equation}
 where $\mathfrak{F}(F_Y,F_Z)$ is the \emph{Fr{\'e}chet class} of all possible joint distributions of $(Y,Z)$ matching the previously defined marginal distributions  $F_{Y}$ and $F_{Z}$. More explicitly, for any joint distribution $F_{YZ} \in \mathfrak{F}(F_Y,F_Z)$ we have $\Pi_{y}\{F_{YZ}\} = F_{Y}$ and $\Pi_{z} \{F_{YZ}\} = F_{Z}$, where
 $\Pi_{.}\{.\}$ denote the projections that take the joint distribution to its (multi-variate) marginals. While we are mainly interested in the application aspect of this, we note that bounds on instrument prices can be derived within the above formulation by taking the risk measure to be the expectation operator.

 It is well known that given a time horizon and confidence level $\alpha$, Value-at-Risk (VaR) is defined as the $\alpha$-percentile of the loss distribution over the specified time horizon. The shortcomings of VaR as a risk measure have been much discussed in the literature. An alternative measure that addresses many of these shortcomings is \emph{Conditional Value at Risk (CVaR)}, also known as \emph{tail VaR} or \emph{Expected Shortfall}. If the loss distribution is continuous, CVaR is the expected loss given that losses exceed VaR. More formally, we have the following definition of CVaR.

  \begin{defn} For the confidence level $\alpha$ and loss random variable $L$, the \emph{Conditional Value at Risk at level $\alpha$} is defined by
   $$\mathrm{CVaR}_{\alpha}(L)=\frac{1}{1-\alpha}\int_{\alpha}^{1}\mathrm{VaR}_{\xi}(L) \, d\xi$$
  \end{defn}

 We will use the following result, which is part of a theorem from~\citet{Schied-2008} (translated into our notation). Here $L$ is regarded as a random variable defined on a probability space $(\Omega,\mathcal{B}, \mathbb{F} )$.
 \begin{theorem}\label{Sup_CVaR_thm}
  $\mathrm{CVaR}_{\alpha}(L)$ can be represented as
   $$\mathrm{CVaR}_{\alpha}=\sup_{G\in \mathcal{G}_{\alpha}} \mathrm{E}_{\mathbb{G}}[L]$$
   where $\mathcal{G}_{\alpha}$ is the set of all probability measures $\mathbb{G} \ll \mathbb{F}$\footnote{$\mathbb{G} \ll \mathbb{F}$ means $\mathbb{G}$ is absolutely continuous with respect to $\mathbb{F}$, i.e. for any $b \in \mathcal{B}$ that $\mathbb{F}(B)=0$ we have $\mathbb{G}(B)=0$. } whose density $d\mathbb{G}/d\mathbb{F}$ is $\mathbb{F}$-a.s. bounded by $1/(1-\alpha)$.
 \end{theorem}
 Applying the above result, with $\rho = \mathrm{CVaR}_{\alpha}$, the worst-case joint distribution problem stated in~(\ref{WCR}) can be conveniently reformulated as:
 \begin{gather}
  \sup_{ F,G\in \mathfrak{F}(F_Y,F_Z) } \mathbb{E}_{G}[L] \\
  \Pi_{y}\{F\} = F_{Y} \nonumber \\
  \Pi_{z}\{F\} = F_{Z} \nonumber \\
  \frac{dG}{dF} \leqslant \frac{1}{1-\alpha} \quad a.s. \nonumber
 \end{gather}
 Note that the final constraint assumes explicitly that the corresponding density exists.

 In many practical cases the marginal distributions will be discrete, either due to a modelling choice, or because they arise from the simulation of separate continuous models for $Y$ and $Z$. In this case, the marginal distributions can be represented by $F_{Y}(Y = y_{m})= p_{m}, m=1,\ldots,M$, and $F_{Z}(Z=z_{n})= q_{n}, \linebreak n=1,\ldots,N$. Any joint distribution of $(Y,Z)$ is then specified by the quantities \linebreak $F_{YZ}(Y=y_{m},Z=z_{n}) = \mathfrak{\psi}_{mn}$, and the worst-case CVaR optimization problem above can be further simplified to:
 \begin{gather}
    \max_{\psi,\mu} \quad \frac{1}{1-\alpha}\sum_{n,m} L_{mn}\cdot\mu_{mn} \label{Sup_CVaR}\\
    \sum_{n} \mathfrak{\psi}_{mn} = p_m \quad m=1,\ldots,M \nonumber \\
    \sum_{m} \mathfrak{\psi}_{mn} = q_n \quad n=1,\ldots,N \nonumber \\
    \sum_{n,m} \mu_{mn}=1-\alpha \nonumber \\
     0  \leqslant \mu_{mn} \leqslant \mathfrak{\psi}_{mm}  \nonumber
 \end{gather}

 Evidently this is a well-defined linear programming problem, and has the general form of a mass transportation problem with multiple constraints. Note that since the sum of each marginal distribution is equal to one, we do not have to include the additional constraint that the total mass of $p$ equals one. Excluding the bounds, this LP has $2mn$ variables and $m+n+1+nm$ constraints. Consequently, the above formulation can lead to very large linear programs. In the numerical examples presented later in this paper, we employ marginal distributions with  market scenarios and credit scenarios ranging from $1,\!000$ to $5,\!000$, yielding joint distributions of $\mathcal{O}(10^7)$. \linebreak Specialized algorithms for linear programs that take advantage of the structure of the transportation problem may well be required for problems defined by marginal distributions with larger numbers of scenarios. The reader if referred to \cite{MTP-Svetlozar-Ruschendorf} for more information on various types of Transportation Problem.

\section{Wrong-way Risk and Counterparty Credit Risk}
The internal ratings based approach in the Basel Accord (\citetalias{BCBS}) provides a formula for the charge for counterparty credit risk capital for a given counterparty that is based on four numerical inputs: the probability of default (PD), exposure at default (EAD), loss given default (LGD) and maturity (M).
 \begin{equation}
    \text{Capital} = \text{EAD} \cdot \text{LGD} \cdot \left[\Phi\left(\frac{\Phi^{-1}(PD)+\sqrt{\rho}\cdot \Phi^{-1}(0.999)}{\sqrt{1-\rho}}\right)\right]\cdot \text{MA(M,PD)}
 \end{equation}
 Here $\Phi$ is the cumulative distribution function of a standard normal random variable, and MA is a maturity adjustment
(see~\citetalias{BCBS}).\footnote{In the most recent version of the charge, exposure at default may be reduced by current CVA, and the maturity adjustment may be omitted,
 if migration is accounted for in the CVA capital charge. See~\citetalias{BCBS} for details.}
 The probability of default is estimated based on an internal rating system, while the LGD is the estimate of a downturn LGD for the counterparty based on an internal model.
 Another parameter appearing in the formula, the correlation ($\rho$), is essentially determined as a function of the probability of default.

 The exposure at default in the above formula is a constant. However, as noted above, counterparty exposures are inherently stochastic in nature, and potentially correlated with counterparty defaults (thus giving rise to wrong-way risk). The Basel accord circumvents this issue by setting $\mathrm{EAD} = \alpha \times \mbox{Effective EPE}$, where Effective EPE is a functional of a given simulation of potential future exposures (see~\citetalias{BCBS}, \citet{DePriscoRosen} or \citet{GarciaCespedes} for detailed discussions). The multiplier $\alpha$ defaults to a value of 1.4; however it can be reduced through the use of internal models (subject to a floor of 1.2). Using internal models, a portfolio's alpha is defined as the ratio of CCR economic capital from a joint simulation of market and credit risk factors ($EC^{Total}$) and the economic capital when counterparty exposures are deterministic and equal to expected positive exposure.\footnote{Expected positive exposure (EPE) is the average of potential future exposure, where averaging is done over time and across all exposure scenarios. See~\citetalias{BCBS}, \citet{DePriscoRosen} or \citet{GarciaCespedes} for detailed expressions of this.}
 \begin{equation}
 \alpha=\frac{EC^{Total}}{EC^{EPE}}
 \end{equation}
 The numerator of $\alpha$ is economic capital based on a full joint simulation of all market and credit risk factors (i.e. exposures are treated as being stochastic, and they are not treated as independent of the credit factors). The denominator is economic capital calculated using the Basel credit model with all counterparty exposures treated as constant and equal to EPE. For infinitely granular portfolios in which PFEs are independent of each other and of default events, one can assume that exposures are deterministic and given by the EPE. Calculating $\alpha$ tells us how far we are from such an ideal case.

 \subsection{Worst-Case Joint Distribution in the Basel Credit Model}
 In this section we demonstrate how the worst-case joint distribution problem can be appliedto the Basel portfolio credit risk model for the purpose of calculating the worst-case alpha multiplier.

 In order to calculate the total portfolio loss, we have to determine whether each of the counterparties in the portfolio has defaulted or not. To do so, we define  the creditworthiness index of each counterparty $k,  1\leqslant k \leqslant K$, using a single factor Gaussian copula as\footnote{In principle we can introduce a fat-tailed copula in place of the Gaussian copula, but  with a considerable increase in the resulting
computational requirement.}:
 \begin{equation} \label{CWI}
 \text{CWI}_k = \sqrt{\rho_k} \cdot Z + \sqrt{1-\rho_k} \cdot \epsilon_k
 \end{equation}
 where $Z$ and $\epsilon_k$ are independent standard normal random variables and $\rho_k$ is the factor loading giving the sensitivity of counterparty $k$ to the systematic factor $Z$. If $\text{PD}_k$ is the default probability of counterparty $k$, then
 that counterparty will default if:
 \begin{equation*}
  \text{CWI}_k \leq  \Phi^{-1}(\text{PD}_k)
  \end{equation*}

 Assuming that we have $M<\infty$ market scenarios in total, if $y_{km}$ is the exposure to counterparty $k$ under market scenario $m$,
  the total loss under each market scenario is:
 \begin{equation} \label{TotalLoss}
 L_m = \sum_{k=1}^{K} y_{km} \cdot \mathbf{1}\left\{ \text{CWI}_k \leq \Phi^{-1}(\text{PD}_k) \right\}
 \end{equation}
 Below we focus on the co-dependence between the market factors $Y$ and the systematic credit factor $Z$. In particular, we assume that the market factors $Y$ and the idiosyncratic credit risk factors $\varepsilon_{k}$ are independent. This amounts to assuming that there is systematic wrong-way risk, but no idiosyncratic wrong-way risk (see~\citet{GarciaCespedes} for a discussion).
 Define the systematic losses under market scenario $m$ to be:
 \begin{equation} \label{SysLoss}
 L_m(Z) = \sum_{k=1}^{K} y_{km} \Phi\left(\frac{\Phi^{-1}(\text{PD}_k) - \sqrt{\rho_k} \cdot Z}{\sqrt{1-\rho_k}}\right)
 \end{equation}
 with probability $\mathbb{P}(Y=y_{m})=p_{m}$. Next we discretize the systematic credit factor $Z$ using $N$ points and define $L_{mn}$ as:
  \begin{equation*}
  L_{mn}(Z) = \sum_{k=1}^{K} y_{km} \Phi\left(\frac{\Phi^{-1}(\text{PD}_k) - \sqrt{\rho_k} \cdot Z_n}{\sqrt{1-\rho_k}}\right)
  \end{equation*}
  \begin{equation*}
 \mathbb{P}(Z=z_n) = q_n \quad \text{for} \quad   1\leq n \leq N
 \end{equation*}
 where $L_{mn}$ represents the losses under market scenario $m$, $1 \leq m \leq M$, and credit scenario $n$, $1 \leq n \leq N$.

 In finding the worst-case joint distribution, we focus on systematic losses, and systematic wrong-way risk, and consequently we need only
 discretize the systematic credit factor $Z$.  We employ a naive discretization of its standard normal marginal:
\begin{equation*}
 \mathbb{P}_{Z}(Z=z_n) = q_n =\Phi(z_n)-\Phi(z_{n-1}) \quad j=1,\ldots,N
\end{equation*}
 where $z_{0} = -\infty$ and $z_{N+1} = \infty$. In the implementation stage in this paper, we set $N= 1000$, and take $z_{j}$ to be equally spaced points in the interval $[-5,5]$. This enables us to consider the entire portfolio loss distribution under the worst-case joint distribution. In calculating risk at a particular confidence level, there is potentially still much scope for improvement over our strategy by choosing a finer discretization of $Z$ in the left tail\footnote{Although we have used an evenly spaced grid for discretizing $Z$, importance sampling techniques can be utilized to better capture the behaviour of worst-case joint distribution in the left tail.}.

 For a given confidence level $\alpha$, the worst-case joint distribution of market and credit factors, $\psi_{mn}, m=1,\ldots,M, n=1,\ldots,N$ can be
 obtained by solving the LP stated in~(\ref{Sup_CVaR}). Having found the discretized worst-case joint distribution,
 we can simulate from the full (not just systematic) credit loss distribution using the following algorithm in order to generate portfolio losses:

 \begin{enumerate}
   \item Simulate a  random market scenario $m$ and credit state $N$ from the discrete worst-case joint distribution $\psi_{mn}$.
   \item Simulate the creditworthiness index of each counterparty. Supposing that $z_{n}$ is the credit state for the systematic
   credit factor from Step 1, simulate $Z$ from the distribution of a standard normal random variable conditioned to be in $(z_{n-1},z_{n})$. Then
   generate $K$ i.i.d. standard normal random variables $\varepsilon_{k}$, and determine the creditworthiness indicators for each
   counterparty using equation~(\ref{CWI}).

   \item Calculate the portfolio loss for the current market/credit scenario: using the above simulated creditworthiness indices and the given default probabilities and
   asset correlations, calculate either systematic credit losses using~(\ref{SysLoss}) or total credit losses using~(\ref{TotalLoss}).
 \end{enumerate}

 \section{Application to Counterparty Credit Risk}
 In this section we consider the use of the worst-case joint distribution problem to calculate an upper bound on the alpha multiplier for counterparty credit risk using a real-world portfolio of a large financial institution. The portfolio consists of over-the-counter derivatives with a wide range of counterparties, and is highly sensitive to many  risk factors, including interest rates and exchange rates. Results calculated using the worst-case joint distribution are compared to those using the stress-testing
 algorithm correlating the systematic credit factor to total portfolio exposure, as described in~\citet{GarciaCespedes} and~\citet{Rosen-Saunders-2009}.
 More specifically, we begin by solving the worst-case CVaR linear program~(\ref{Sup_CVaR}) for a given, pre-computed set of exposure scenarios,\footnote{Exposures
 are single-step EPEs based on a multi-step simulation using a model that assumes mean reversion for the underlying stochastic factors.}
 and the discretization of the (systematic) credit factor in the single factor Gaussian copula credit model described above. We then simulate the full model
 based on the resulting joint distribution, under the assumption of no idiosyncratic wrong-way risk (so that the market factors and the idiosyncratic
 credit risk factors remain independent of each other).

 The market scenarios are derived from a standard Monte-Carlo simulation of portfolio exposures, so that we have:
 \begin{equation}
 \mathbb{P}_{Y}(Y=y_m) = p_m =\frac{1}{M} \quad i=1,\ldots,M
 \end{equation}

 In the coming section we will look at top counterparties with respect to total portfolio exposure and some of their exposure characteristics.

 \subsection{Portfolio Characteristics}
 The analysis presented in this section is based on a large portfolio of over-the-counter derivatives including positions in interest rate swaps and credit default swaps with approximately 4,800 counterparties. We focus on two cases, the largest  220 and largest 410 counterparties as ranked by exposure (EPE); these two cases account for more than $95\%$ and $99\%$ of total portfolio exposure respectively.

\begin{figure}[!h]
 \begin{center}
 \vspace*{-4.5cm}
 \hspace*{1.0cm}\scalebox{0.5}{\includegraphics{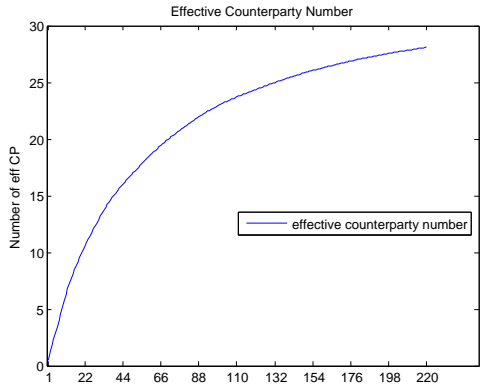}}
 \vspace*{-4.5cm}
  \end{center}
  \caption[c]{Effective number of counterparties for the largest 220 counterparties. }\label{EffCP220}
 \end{figure}

\begin{figure}[!ht]
 \begin{center}
 \vspace*{-4.5cm}
 \hspace*{1.0cm}\scalebox{0.5}{\includegraphics{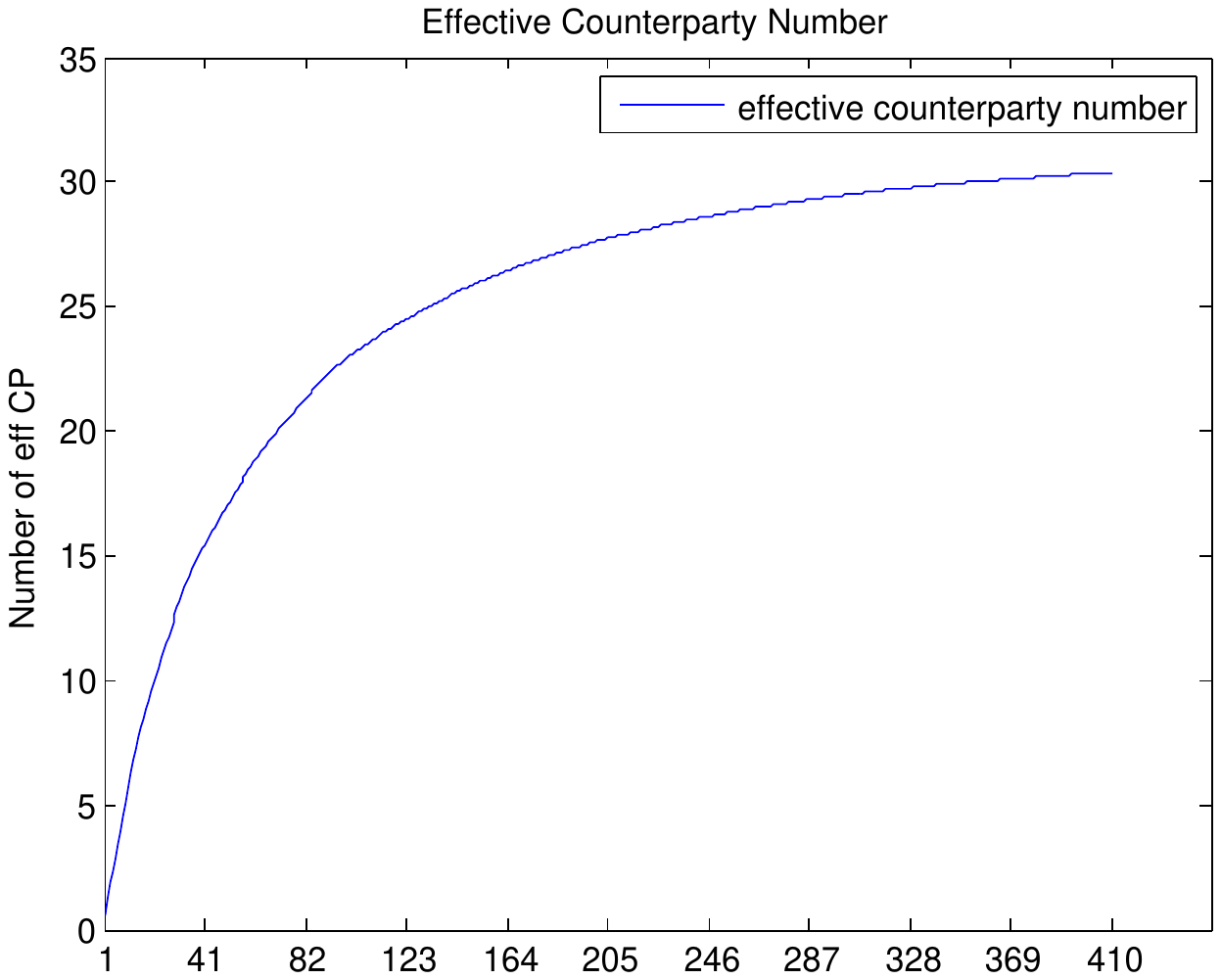}}
 \vspace*{-4.5cm}
 \caption{  Effective number of counterparties for the largest 410 counterparties. }\label{EffCP410}
 \end{center}
 \end{figure}

 Figures~\ref{EffCP220} and~\ref{EffCP410} present exposure concentration reports, giving the number of effective counterparties among the largest 220 and 410 counterparties
 respectively.\footnote{Counterparty exposures (EPEs) are sorted in decreasing order.
 Let $w_n$ be the $n^{th}$ largest exposure; then the Herfindahl index of the $N$ largest exposures is defined as:
 $$H_N=\nicefrac{\sum_{n=1}^{N}w_n^2}{\left(\sum_{n=1}^{N}w_n\right)^2}$$
 The effective number of counterparties among the $N$ largest counterparties with respect to total portfolio exposures is $H_{N}^{-1}$.}
 The effective number of counterparties for the entire portfolio in shown in Figure $3$. As can be seen in these figures the choice of largest $220$ and $410$
 counterparties is justified as the number of effective counterparties for the entire portfolio is 31 in each case.

 \begin{figure}[!ht]
 \begin{center}
 \vspace*{-4.5cm}
 \hspace*{1.0cm}\scalebox{0.5}{\includegraphics{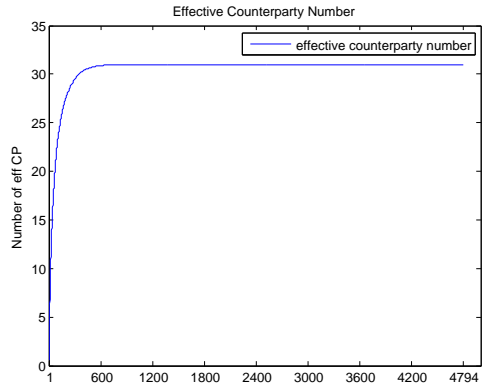}}
 \vspace*{-4.5cm}
 \caption{  Effective number of counterparties for the entire portfolio. }
 \end{center}
 \end{figure}

 The exposure simulation uses $M=1000$ and $M=2000$ market scenarios, while the systematic credit risk factor is discretized with $N=1000$, $N=2000$ and $N=5000$ using the
 method described above. For CVaR calculations, we employ the 95\% and 99\% confidence levels; This has been dictated by our choice of $N$. As we mentioned earlier, importance sampling methods would be more suitable at higher confidence levels. Note that there would be a dramatic increase in the computational cost of simulating the number of exposure scenarios in this case.

\begin{figure}[!h]
  \begin{center}
  \vspace*{-4.5cm}
  \hspace*{1.0cm}\scalebox{0.5}{\includegraphics{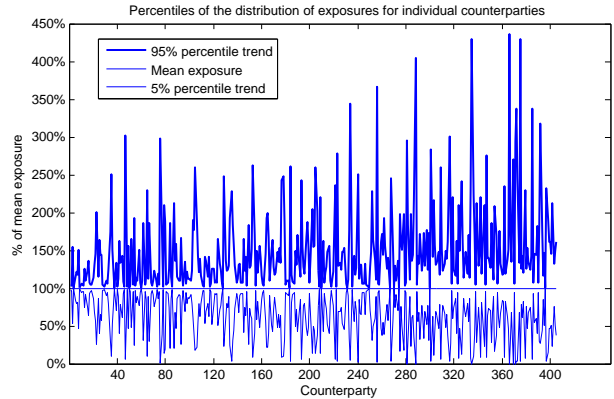}}
  \vspace*{-4.5cm}
  \caption{5\% and 95\% percentiles of the exposure distributions of individual counterparties, expressed as a
  percentage of  counterparty mean exposure (Here counterparties are sorted in order of decreasing mean exposure).   }\label{Perc_Expo}
  \end{center}
  \end{figure}

\begin{figure}[!h]
  \begin{center}
  \vspace*{-4.5cm}
  \hspace*{1.0cm}\scalebox{0.5}{\includegraphics{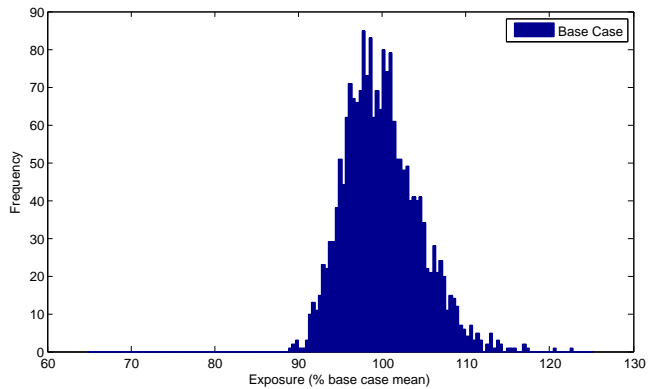}}
  \vspace*{-4.5cm}
  \caption{Histogram of total portfolio exposures from the exposure simulation.}\label{BaseCaseExpo}
  \end{center}
  \end{figure}

  The ranges of individual counterparty exposures are plotted in figure \ref{Perc_Expo}. The $95^{th}$ and $5^{th}$ percentiles of the exposure distribution are given as a percentage of the mean exposure for each counterparty. The volatility of the counterparty exposure tends to increase as the the mean exposure of the respective counterparties decreases. In other words, counterparties with higher mean exposure tend to be less volatile compared to counterparties with lower mean exposure. Given the above characteristics, we would expect that wrong-way risk could have an important impact on portfolio risk, and that the contribution of idiosyncratic risk will also be significant.
  The distribution of the total portfolio exposures from the exposure simulation is given in Figure~\ref{BaseCaseExpo}. The histogram shows that the portfolio exposure distribution is both leptokurtic and highly skewed. It is important to employ such highly skewed with very fat tail exposure distribution to ensure the proper conservatism of our method in practice.

\subsection{Numerical Results}
  To assess the severity of the worst-case joint distribution, and to determine the degree of conservativeness in earlier methods, we compare
  risk measures calculated using the worst-case joint distribution to those computed based on the stress-testing algorithm presented in~\citet{GarciaCespedes}
  and~\citet{Rosen-Saunders-2009}. In this method, exposure scenarios are sorted in an economically meaningful way, and then a two-dimensional copula
  is applied to simulate the joint distribution of exposures (from the discrete distribution defined by the exposure scenarios) and the systematic credit factor. The
  algorithm is efficient, and preserves the (simulated) joint distribution of the exposures. Here, we apply a Gaussian copula, and sort exposure scenarios by the
  value of total portfolio exposure (this intuitive sorting method has proved to be conservative in many tests, conducted in~\citet{Rosen-Saunders-2009}).
  For each level of correlation in the Gaussian copula, we calculate the ratio of risk (as measured by 95\% and 99\% CVaR) estimated using the sorting method to risk, which in turn, is estimated using the worst-case loss distribution.

  We present the results of three discretizations of the worst-case joint distribution. Case I employs $M=1,\!000$ market scenarios and $N=1,\!000$ credit scenarios; Case II doubles the market and credit scenarios. Lastly in Case III we use $2,\!000$ market scenarios and $5,\!000$ credit scenarios. Note that Case I and Case II yield a discretized worst-case distribution of $\mathcal{O}(10^6)$ while Case III's output is of $\mathcal{O}(10^7)$.

{\scriptsize \begin{table} \captionsetup{font=scriptsize} {\scriptsize \begin{center} \begin{tabular}{lll}
\hlinewd{0.8pt}
 \textbf{Case I} &  &  \\
 \textbf{$MN= \mathcal{O} (10^6)$} & $M = 1000 \quad \text{market  scenarios}$ & $N = 1000 \quad\text{credit scenarios}$ \\ \hline
 $\alpha$ & $\min (\text{CVaR}_\text{sys} / \text{CVaR}_\text{wcc}) $ & $\max (\text{CVaR}_\text{sys} / \text{CVaR}_\text{wcc}) $ \\ \hline
 0.95   & 52.2\%                            & 95.1\%  \\
 0.99   & 51.8\%                            & 95.9\%   \\ \hline
  \textbf{Case II} &  &  \\
 \textbf{$MN= \mathcal{O} (10^6)$} & $M = 2000 \quad\text{market scenarios}$ & $N = 2000 \quad\text{credit scenarios}$ \\ \hline
 $\alpha$ & $\min (\text{CVaR}_\text{sys} / \text{CVaR}_\text{wcc}) $ & $\max (\text{CVaR}_\text{sys} / \text{CVaR}_\text{wcc}) $ \\ \hline
 0.95   & 50.3\%                            & 96.9\%  \\
 0.99   & 49.6\%                            & 96.6\%   \\ \hline
   \textbf{Case III} &  &  \\
 \textbf{$MN= \mathcal{O} (10^7)$} & $M = 2000 \quad\text{market scenarios}$ & $N = 5000 \quad\text{credit scenarios}$ \\ \hline
 $\alpha$ & $\min (\text{CVaR}_\text{sys} / \text{CVaR}_\text{wcc}) $ & $\max (\text{CVaR}_\text{sys} / \text{CVaR}_\text{wcc}) $ \\ \hline
 0.95   & 44.8\%                            & 96.4\%  \\
 0.99   & 44.1\%                            & 97.1\%   \\
 \hlinewd{0.8pt}
\end{tabular} \end{center}
\caption{Minimum and maximum of ratio of systematic CVaR using the Gaussian copula algorithm to systematic CVaR using the worst-case joint distribution for the largest 220 counterparties at 95\% and 99\% confidence level.} \label{CVaR_220_table}
} \end{table} }

\begin{figure}
\begin{tabular}{cc}\captionsetup{font=scriptsize}
  \includegraphics[width=65mm]{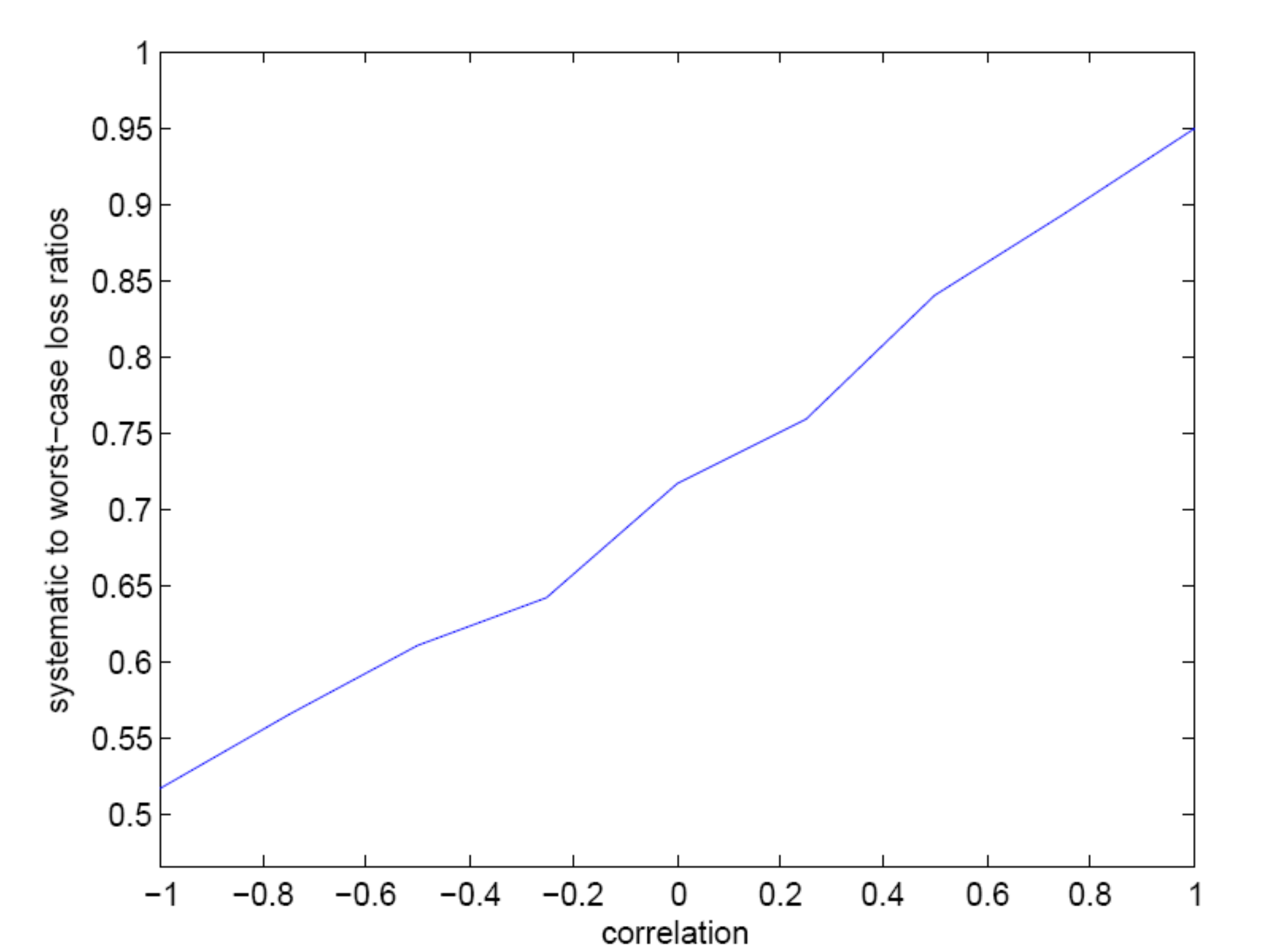} &   \includegraphics[width=65mm]{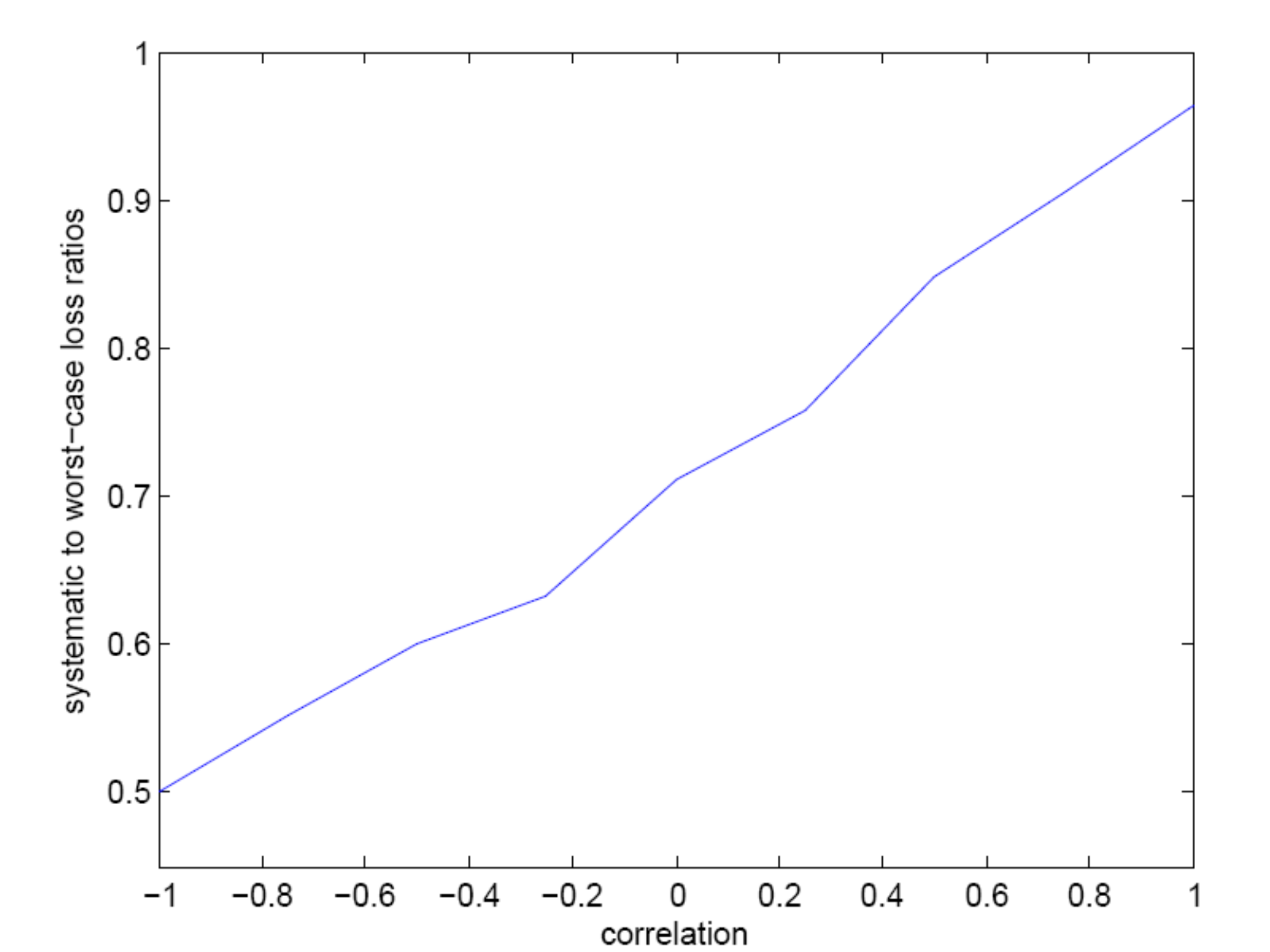} \\
Case I & Case II \\[6pt]
\multicolumn{2}{c}{\includegraphics[width=65mm]{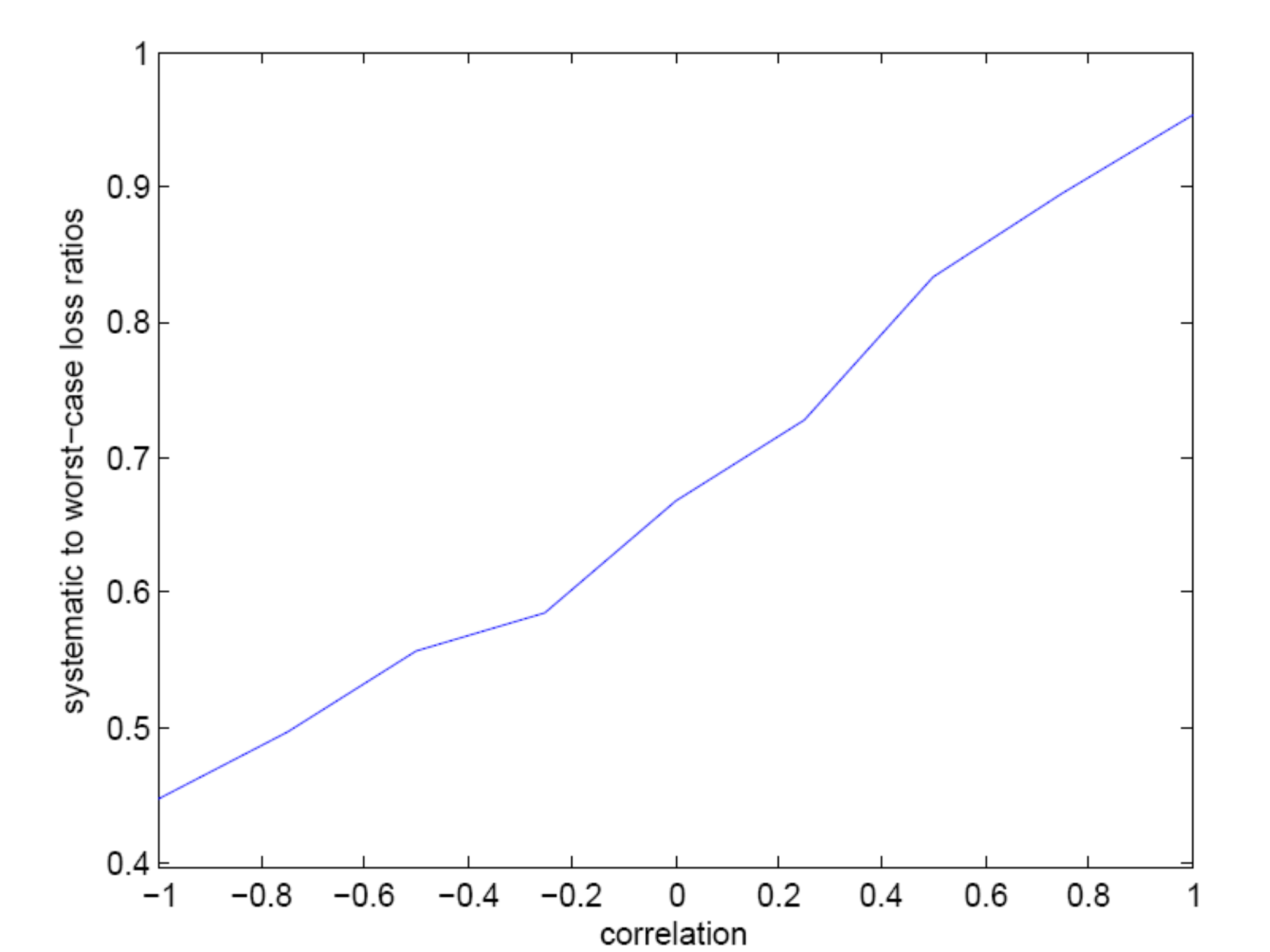} }\\
\multicolumn{2}{c}{Case III}
\end{tabular}
\caption{Ratio of systematic CVaR using the Gaussian copula algorithm to systematic CVaR using the worst-case joint distribution for the largest 220 counterparties at 95\% confidence level.}\label{LR_sys_220}
\end{figure}

{\scriptsize \begin{table}  \captionsetup{font=scriptsize} {\scriptsize \begin{center} \begin{tabular}{lll}
\hlinewd{0.8pt}
 \textbf{Case I} &  &  \\
 \textbf{$MN= \mathcal{O} (10^6)$} & $M = 1000 \quad \text{market  scenarios}$ & $N = 1000 \quad\text{credit scenarios}$ \\ \hline
 $\alpha$ & $\min (\text{CVaR}_\text{tot} / \text{CVaR}_\text{wcc}) $ & $\max (\text{CVaR}_\text{tot} / \text{CVaR}_\text{wcc}) $ \\ \hline
 0.95   & 52.2\%                            & 96.8\%  \\
 0.99   & 51.6\%                              & 96.5\%   \\ \hline
  \textbf{Case II} &  &  \\
 \textbf{$MN= \mathcal{O} (10^6)$} & $M = 2000 \quad\text{market scenarios}$ & $N = 2000 \quad\text{credit scenarios}$ \\ \hline
 $\alpha$ & $\min (\text{CVaR}_\text{tot} / \text{CVaR}_\text{wcc}) $ & $\max (\text{CVaR}_\text{tot} / \text{CVaR}_\text{wcc}) $ \\ \hline
 0.95   & 49.6\%                            & 97.2\%  \\
 0.99   & 48.9\%                              & 97.4\%   \\ \hline
  \textbf{Case III} &  &  \\
 \textbf{$MN= \mathcal{O} (10^7)$} & $M = 2000 \quad\text{market scenarios}$ & $N = 5000 \quad\text{credit scenarios}$ \\ \hline
 $\alpha$ & $\min (\text{CVaR}_\text{tot} / \text{CVaR}_\text{wcc}) $ & $\max (\text{CVaR}_\text{tot} / \text{CVaR}_\text{wcc}) $ \\ \hline
 0.95   & 45.8\%                            & 98.1\%  \\
 0.99   & 44.7\%                                & 98.2\%   \\
 \hlinewd{0.8pt}
\end{tabular} \end{center}
\caption{Minimum and maximum ratio of CVaR for total losses using the Gaussian copula algorithm to CVaR for total losses using the worst-case joint distribution for the largest 410 counterparties at 95\% and 99\% confidence level.} \label{CVaR_410_table}
} \end{table} }

\begin{figure}
\begin{tabular}{cc}\captionsetup{font=small}
  \includegraphics[width=65mm]{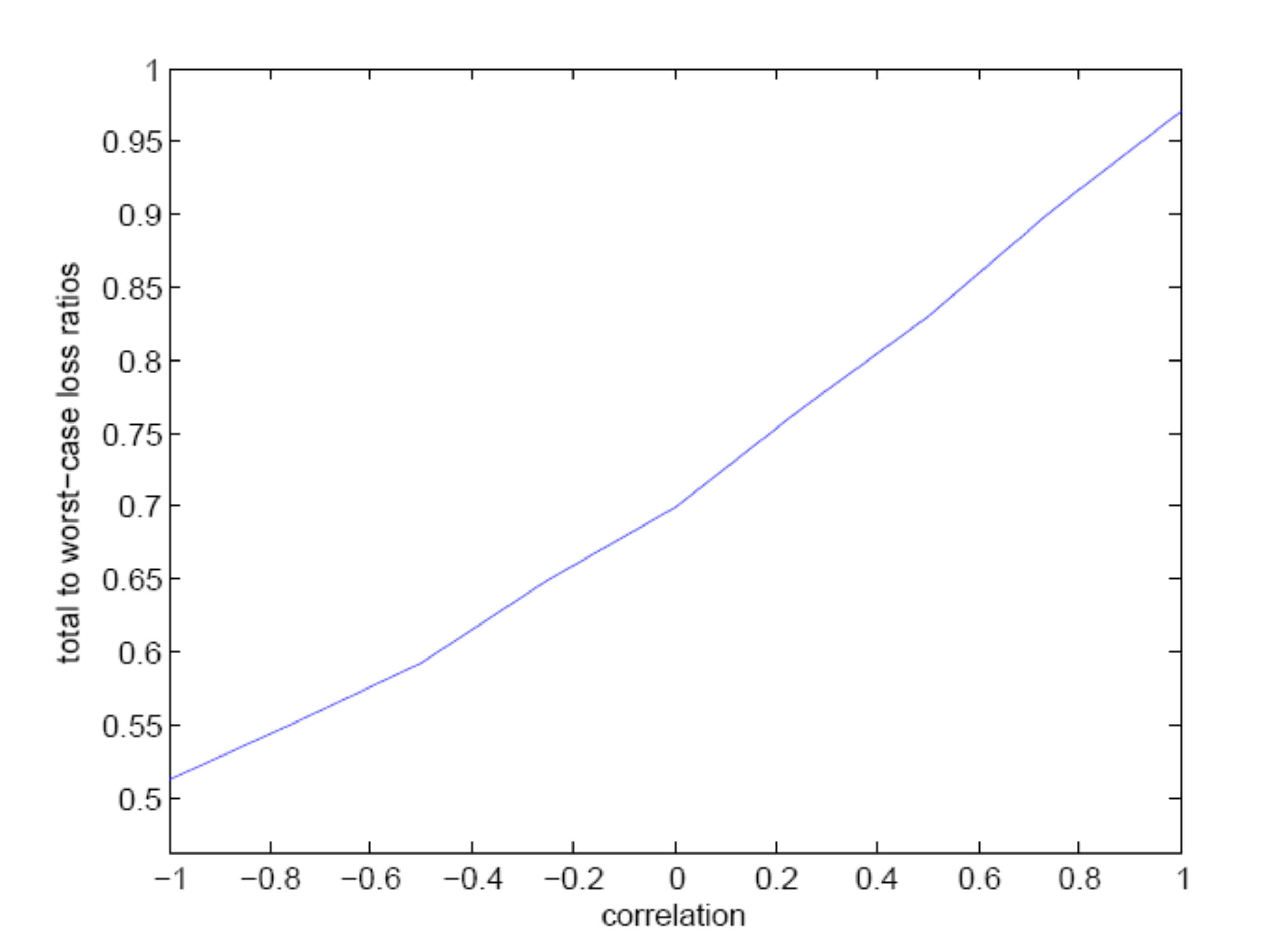} &   \includegraphics[width=65mm]{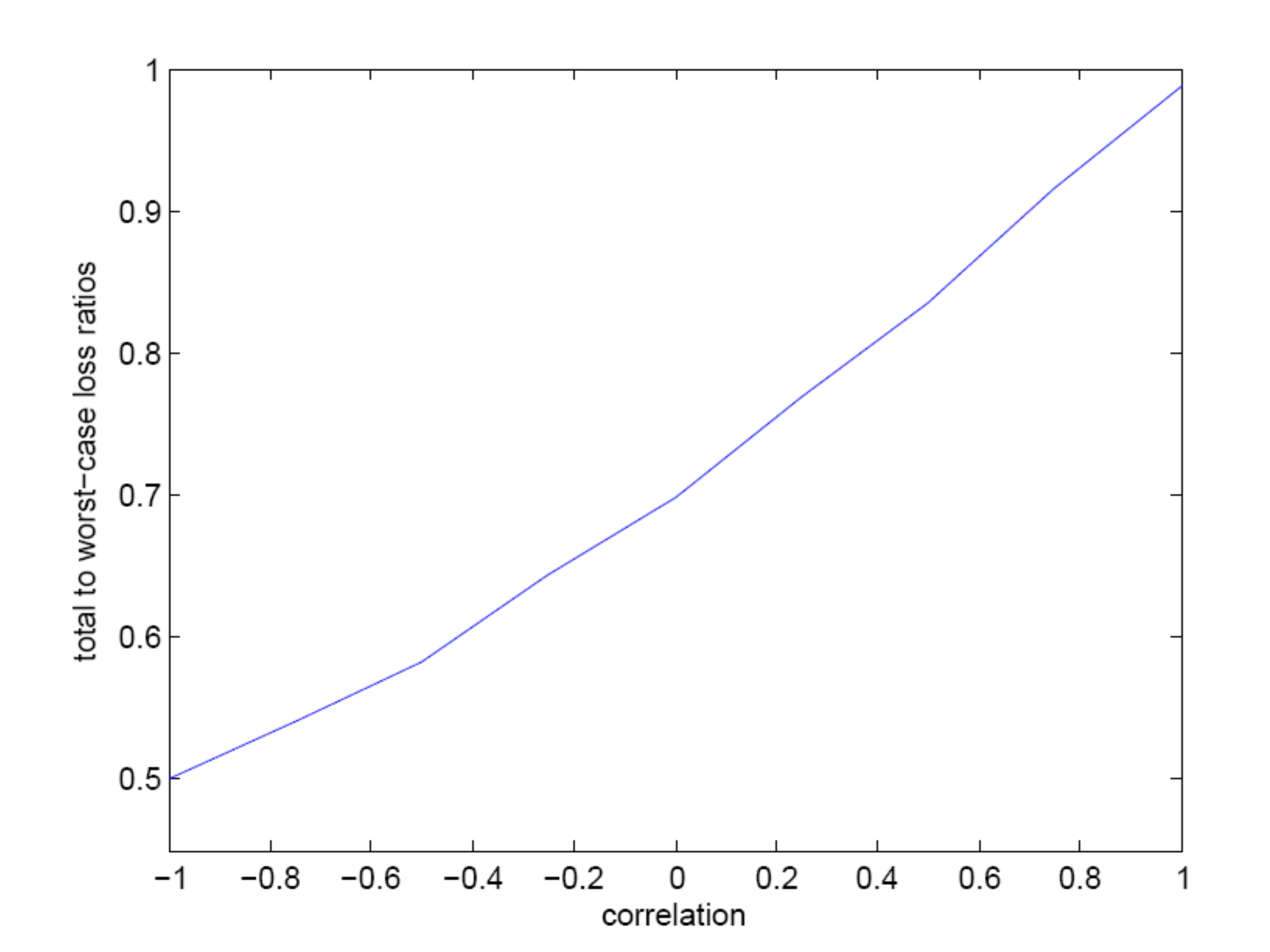} \\
 Case I & Case II \\[6pt]
\multicolumn{2}{c}{\includegraphics[width=65mm]{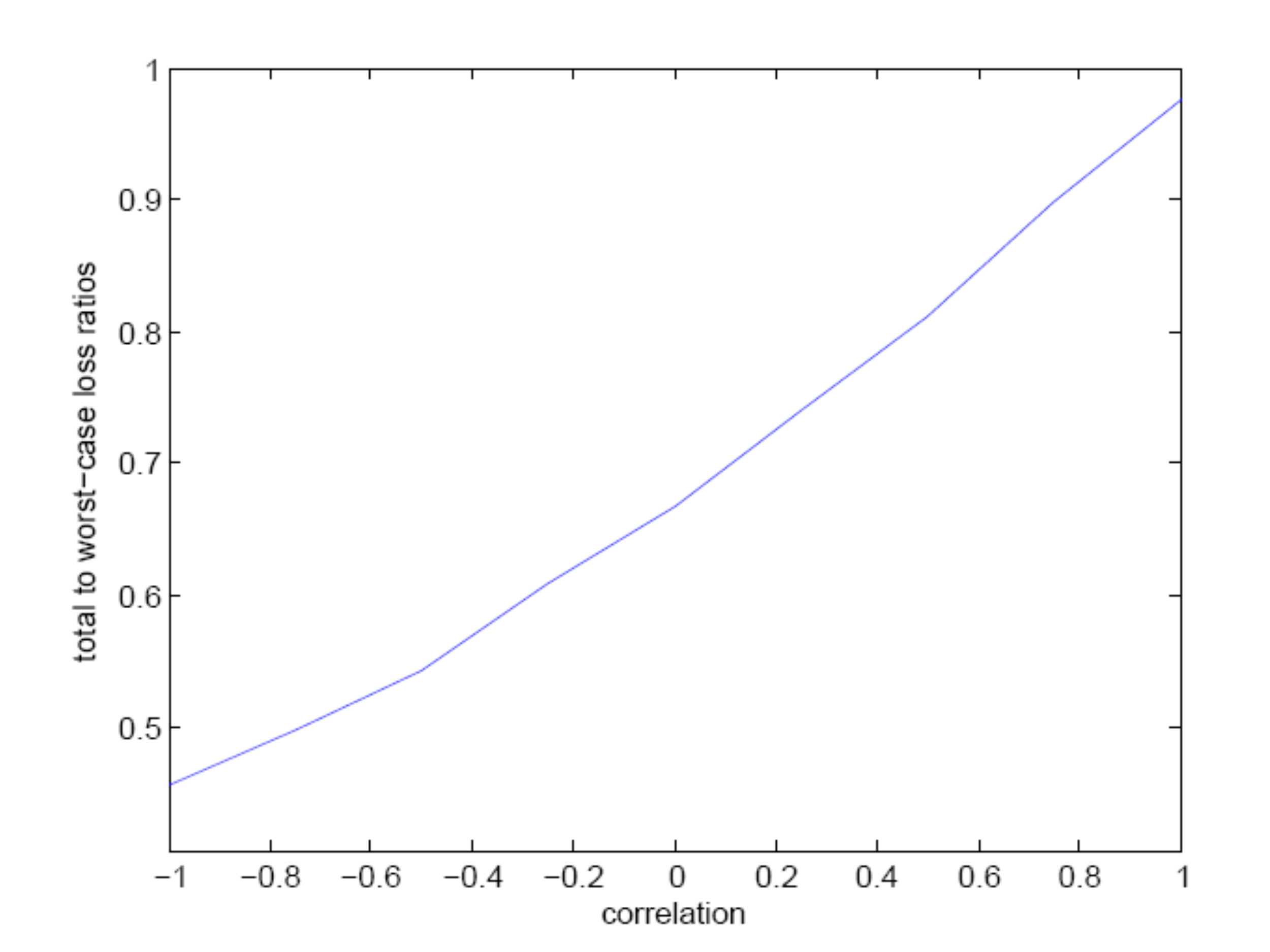} }\\
\multicolumn{2}{c}{Case III}
\end{tabular}
\caption{Ratio of CVaR for total losses using the Gaussian copula algorithm to CVaR for total losses using the worst-case joint distribution for the largest 410 counterparties at 99\% confidence level.} \label{LR_tot_410}
\end{figure}

 Figure~\ref{LR_sys_220} shows the graph of the ratio of systematic CVaR using the Gaussian copula algorithm described in \citet{Rosen-Saunders-2009} to systematic CVaR using the worst-case joint distribution method described in section 3 for the largest 220 counterparties at 95\% confidence level. The ratio of the CVaR of the systematic portfolio loss to the CVaR calculated using worst-case joint distribution across various levels of market-credit correlation and discretization scenarios indicate that the distribution simulated using the worst-case joint distribution has a higher CVaR compared to previous simulation methods for the largest 220 counterparties by $4.9\%$, $3.1\%$ and $3.6\%$ at $\alpha=0.95$ respectively when the systematic risk factor and market risk factor are fully correlated. The difference is larger for lower levels of market-credit correlation in the stress testing algorithm. The sorting methods do indeed produce relatively conservative numbers (at high levels of market-credit correlation) for this portfolio. In addition to the results presented in figure~\ref{LR_sys_220}, table~\ref{CVaR_220_table} shows the minimum and maximum of systematic CVaR to CVaR calculated from worst-case joint distribution at 99\% confidence level. Note that the results are consistent with what we observed at $\alpha=95\%$.

 Similar results for calculating CVaR ratios using the total portfolio loss for the largest 410 counterparties which constitute more than $99.6\%$ of total portfolio exposure are shown in figure~\ref{LR_tot_410}. The graphs are based on a higher confidence level, $\alpha=99\%$, compared to figure~\ref{LR_sys_220}. Table~\ref{CVaR_410_table} shows comparable results to those presented in table~\ref{CVaR_220_table} when we use total portfolio loss instead of systematic loss.

\section{Conclusion and Future Work}
In this paper, we studied the problem of finding the worst-case joint distribution of a set of risk factors given prescribed multivariate marginals and a nonlinear
loss function. We showed that when the risk measure is CVaR, and the distributions are discretized, the problem can be solved conveniently using linear programming.
The method has applications to any situation where marginals are provided, and bounds need to be determined on total portfolio risk. This arises in many
financial contexts, including pricing and risk management of exotic options, analysis of structured finance instruments, and aggregation of portfolio risk across risk types.
Applications to counterparty credit risk were emphasized in this paper, and they include assessing wrong-way risk in the credit valuation adjustment, and counterparty
credit risk measurement. A detailed application of the algorithm for counterparty risk measurement on a real portfolio was subsequently presented and discussed.

 The method presented in this paper will be of interest to regulators, who are interested in determining how conservative dependence structures estimated (or assumed) by risk managers in industry actually are. It will also be of interest to risk managers, who can employ it to stress test their assumptions regarding dependence in risk measurement calculations.

\bibliographystyle{plainnat}
\bibliography{Ref}

\end{document}